\documentclass[alpha-refs]{nbdt-article}

\usepackage{siunitx}

\papertype{Original Article}

\paperfield{Commentary}

\title{Appreciating the variety of goals in computational neuroscience}

\author[1]{Konrad P.~Kording PhD}
\author[2]{Gunnar Blohm PhD}
\author[3]{Paul Schrater PhD}
\author[4]{Kendrick Kay PhD}

\affil[1]{Departments of Bioengineering and Neuroscience, University of Pennsylvania, Philadelphia, PA, USA}
\affil[2]{ViKinG Lab, Centre for Neuroscience Studies, Queen's University, Kingston, ON, Canada}
\affil[3]{Departments of Psychology and Computer Science, University of Minnesota, Minneapolis, MN, USA}
\affil[4]{Center for Magnetic Resonance Research, Department of Radiology, University of Minnesota, Minneapolis, MN, USA}

\corraddress{Kendrick Kay PhD, Center for Magnetic Resonance Research, Department of Radiology, University of Minnesota, Minneapolis, MN, USA}
\corremail{kendrickkay@gmail.com}

\fundinginfo{G.B. was supported by the National Science and Engineering Research Council (NSERC, Canada).}

\runningauthor{Konrad Kording et al.}

\begin{document}

\maketitle

\begin{abstract}
Within computational neuroscience, informal interactions with modelers often reveal wildly divergent goals. In this opinion piece, we explicitly address the diversity of goals that motivate and ultimately influence modeling efforts. We argue that a wide range of goals can be meaningfully taken to be of highest importance. A simple informal survey conducted on the Internet confirmed the diversity of goals in the community. However, different priorities or preferences of individual researchers can lead to divergent model evaluation criteria. We propose that many disagreements in evaluating the merit of computational research stem from differences in goals and not from the mechanics of constructing, describing, and validating models. We suggest that authors state explicitly their goals when proposing models so that others can judge the quality of the research with respect to its stated goals.

\end{abstract}


\section{Introduction}

\subsection{Diversity of modeling goals}

Models are essential for progress in neuroscience and exist in a variety of forms and flavors. Here we will follow the definition of `computational model' in Merriam-Webster Dictionary \citep{websterMerriamWebsterDictionaryInternational2016}: ``a system of postulates, data, and inferences presented as a mathematical description of an entity or state of affairs''. Within neuroscience, such models come in many flavors. Models can summarize existing data. They can jointly describe brain and behavior data. They can express relations that can be tested in experiments and predict successful clinical treatments. Formulating a concrete model can help uncover hidden assumptions and help assess the suitability of hypothesized relationships. Formulating models can provide mathematical insights and simulating them can lead to systems that solve real-world problems. Accordingly, there is a large community of neuroscientists who construct and use models.

As computational neuroscientists, we became interested in the goals of modeling when we noticed stark differences across models in different papers and fields of neuroscience. For example, when Kendrick studies nonlinearities in the human brain, he cares most about macroscopic measurements and model interpretability \citep{kayBottomupTopdownComputations2017a}. When Paul studies representations decoded from the brain, he cares most about interpretability and representations \citep{carlsonPatternsActivityCategorical2003a}. When Gunnar writes a paper about linear-systems explanations of eye movements, he cares most about behavior, mathematical simplicity, and the real-world relevance of the task \citep{orbandexivryKalmanFilteringNaturally2013}. One might suspect that these differences in goals stem merely from differences in modeling methodology. However, when Konrad writes a paper using the same methodology as Gunnar (i.e. linear systems), he cares most about the model being the optimal solution to a computational problem \citep{kordingDynamicsMemoryConsequence2007}.  Thus, there appears to be a diversity of modeling goals with real impact on the way we organize our research.

Despite this diversity, outsiders often perceive computational neuroscience as being homogeneous. What unites computational neuroscience is a commitment to an approach that combines mathematical reasoning with computer simulations. However, this approach is applied across a broad array of topics and within each topic, researchers strive to achieve distinct goals. At Society for Neuroscience, computational approaches are often corralled out of ease into a single section despite differing goals. When experimentalists add computation to papers or grants, they often do so without choosing a goal first. Young scientists declare they want to do computation without first committing to a goal. And lastly, when neuroscientists (admittedly insiders) write books, they tend to merge computational approaches, despite vastly differing goals. This creates the false illusion of homogeneity of a field, whereas we believe that computational neuroscience is, rather, the accumulation of the computational branches of many different fields.

The diversity of goals within computational neuroscience is not without consequence: we propose that many disagreements in evaluating the merit of computational research stem from differences in goals and not from the mechanics of constructing, describing, and validating models.  Goals affect the way science is reviewed. They form key criteria \citep{blohmHowtomodelGuideNeuroscience2018,schraterModelingNeuroscienceDecision2019} that inform both reviewers' and editors' decisions.  Goals are implicitly invoked when consuming and evaluating research, and therefore impact an article's likelihood of success.  Across several disciplines, both meta-analyses and editorial comment \citep{bornmannContentAnalysisReferees2010,byrneCommonReasonsRejecting2000,piersonTop10Reasons2004,throwerEightReasonsRejected2012} provide evidence that editors' and reviewers' preferred goals are criteria to which authors must conform for success\footnote{Examples of these preferences are not hard to find. For example, ``Research doesn't add value to the journal. Sometimes the findings of a research aren't appealing to the journals, especially if those findings do not really contribute to any advancement in their field. If this is the case, it's likely that the paper would be rejected.'' \citep{mukherjee11ReasonsWhy2018}, and ``It's boring. ... The question behind the work is not of interest in the field. The work is not of interest to the readers of the specific journals.'' \citep{throwerEightReasonsRejected2012}. Editors reject theory papers if they do not directly explain empirical data. For example, at \emph{PLoS Computational Biology}, the criterion ``Significant biological insight and general interest to life scientists'' often excludes theory papers that do not prioritize biological realism.}. In our considerable experience as editors, we find that disagreement regarding what constitutes a worthwhile goal for modeling is one of the main drivers of paper rejections.  We believe the problem is that editors' and reviewers' preferred goals are implicit.  By making modeling goals explicit, authors, reviewers, and editors can start to find common ground for the merits of a paper.  

\subsection{A short list of modeling goals}

To the extent that the goals we choose for modeling matter, an important open question is: what exactly are these goals? Examining a broad range of papers in computational neuroscience, we gleaned a variety of different modeling goals, typically revealed in the Introduction, Methods, and Discussion sections \citep{blohmHowtomodelGuideNeuroscience2018}. While it is impossible to produce an exhaustive list, we compile here a list of the most salient and common ones \citep{schraterModelingNeuroscienceDecision2019}.

\begin{itemize}

\item \textbf{Useful} (can be applied to other domains). Some models of the nervous system are also good at solving real-world problems. Models can be evaluated in terms of how good they are at solving such problems. For example, a model of the visual system might be able to solve challenging problems in computer vision \citep{fukushimaNeocognitronSelfOrganizing1980a,serreRealisticModelingSimple2004}. This assumes that the modeled system in the brain is solving a problem that also appears in technical systems.

\item \textbf{Normative} (best possible given certain assumptions). Some models provide the optimal solutions to problems that exist in the real world \citep{chaterRationalAnalysisMind2000,chaterTenYearsRational1999,knillPerceptionBayesianInference1996,todorovOptimalFeedbackControl2002}. Models can be evaluated in terms of how well they represent an optimal solution to a meaningful problem. Normative models are thus often used in domains where behavior or neural properties are expected to be optimal or near optimal \citep{acunaStructureLearningHuman2010,dayanTheoreticalNeuroscienceComputational2001,kordingDecisionTheoryWhat2007}. For example, a model may ask how well people minimize energy when walking \citep{selingerHumansCanContinuously2015}. Thus, we might ask whether a model supplies the optimal solution to a computational problem faced by the brain and how similar behavior is to these predictions. A normative model can also ask whether the assumed principles underlying the optimality criterion are biologically accurate. This assumes that we can understand the goals of a system and that we gain insight if a system appears to optimize what it is expected to optimize \citep{barlowPossiblePrinciplesUnderlying1961,mayrWhatMakesBiology2004}.

\item \textbf{Clinically relevant} (helps healthcare). Some models produce insights that are relevant for developing or evaluating clinical interventions. Models may be evaluated in terms of how well they generalize to medical problems. For example, simulating individual differences with respect to electrical stimulation enables us to place electrodes to maximize stimulation outcome \citep{baiComputationalComparisonConventional2019}. Given the potential to reduce human suffering, there is no doubt that clinical relevance is a meaningful goal. In order for modeling insights to transfer to medicine, a model must be sufficiently similar to the real system.

\item \textbf{Inspire experiments} (untested assumptions, new hypotheses). Some models change the way we think about a problem and thereby raise interesting new hypotheses via abductive inference \citep{josephsonAbductiveInferenceComputation1996,lombrozoExplanationAbductiveInference2012}. Models can be evaluated in terms of the richness of potential experiments they inspire. For example, a model may  suggest that spike timing may affect plasticity and therefore lead to a broad set of tests \citep{danSpikeTimingdependentPlasticity2004,gerstnerNeuronalLearningRule1996}. A formal model might also uncover hidden assumptions that a field makes when considering a proposed mechanism. To inspire experiments, a set of potential models must be small enough such that experimental tests are meaningful.

\item \textbf{Microscopic realism} (looks like the brain). Some models describe the microscopic properties of the brain, such as synaptic, pharmacological, and cellular-level properties. Models can then be evaluated in terms of how well they quantitatively describe those properties. For example, models may predict changes in synapses over time \citep{zadorBiophysicalModelHebbian1990}. Commitment to microscopic realism assumes that microscopic properties can be sufficiently decoupled from macroscopic properties such that a reductionist understanding of neural properties is possible \citep{gillettReductionEmergenceScience2016}.

\item \textbf{Macroscopic realism} (looks like the brain at the population level). Some models describe properties of brain areas and networks.  Models can then be evaluated in terms of how well they quantitatively describe those properties. For example, models may predict the population activity of brain areas  as measured by EEG \citep{al-nashashEEGSignalModeling2004}. Commitment to macroscopic realism assumes that macroscopic properties can be sufficiently decoupled from finer-scale, distributed properties \citep{bennettPhilosophicalFoundationsNeuroscience2003}.

\item \textbf{Behavioral realism} (looks like real behavior). Some models can faithfully describe and explain behavioral phenomena. Models can then be evaluated in terms of how well they quantitatively account for behavior. For example, models can predict the way we move our arm as a function of distance we need to travel \citep{harrisSignaldependentNoiseDetermines1998}. An approach based on behavioral realism supposes that behavior can be understood without a deeper understanding of the brain and that compact models of behavior are possible \citep{greenAlterationsChoiceBehavior2010,krakauerNeuroscienceNeedsBehavior2017,taoCorrectiveResponseTimes2018}.

\item \textbf{Representational} (codes like the brain). Some models aim to use representations of information that are similar to representations in the brain. Models can then be evaluated in terms of how well they quantitatively describe representations. For example, models predict that neurons in motor cortex have cosine tuning \citep{olshausenSparseCodingSensory2004}. Such modeling assumes that representations can be compactly understood and are the basis of the phenomena we want to understand \citep{churchlandNeuralRepresentationNeural1990}.

\item \textbf{Compact} (few short equations). Some models can be succinctly expressed in mathematical language and/or computer code \citep{burgessOccamRazorScientific1998,liIntroductionKolmogorovComplexity2019}. Models can then be evaluated in terms of how well they trade off complexity against the quality of description of the phenomena. For example, Fitt's law can compactly describe the balance between speed and precision during hand movements \citep{fittsInformationCapacityDiscrete1966}. This approach assumes that the phenomenon of interest has a low-complexity description \citep{burgessOccamRazorScientific1998}.

\item \textbf{Analytically tractable} (exact solutions exist). Some models are understandable through mathematical equations as opposed to numerical simulations. Models can then be evaluated in terms of how well they can be analytically solved. For example, models may allow the combination of cues with neurally realistic properties while being analytically solvable \citep{maBayesianInferenceProbabilistic2006a}. For scientists with mathematical training, an analytic approach provides a more generalizable understanding compared to numerical models. An implicit assumption is that the system of interest is sufficiently similar to the analytically tractable model such that analyzing one provides insights into the other \citep{parkerComputerSimulationPhilosophy2012}.

\item \textbf{Interpretable} (relates directly to something the brain does). Some models are easily interpreted with respect to how they work (e.g. what outcomes they predict) and/or how the brain might implement the computations. Models can then be evaluated in terms of how well humans can interpret their meaning. For example, units in a simulated system may have receptive fields similar to those of real neurons \citep{blohmDecodingCorticalTransformations2009,olshausenSparseCodingSensory2004}. For many scientists, prioritizing the interpretability of a model makes the model more relatable to their way of thinking about the brain.

\item \textbf{Beauty} (elegant). Some models may be symmetrical, balanced, or resonate well with the way we think. Models can then be evaluated in terms of how well they resonate intuitively with their target audience. For example, the same model can be presented in the languages of physics, math, and biology, and can be distinctly useful for these different communities \citep{chandrasekharTruthBeautyAesthetics2013,russellMysticismLogicOther2019}.

\end{itemize}

\section{Methods}

To assess modeling goals in the computational neuroscience community, we constructed an online survey using Google Forms. Each of the authors then contacted colleagues via personal e-mails, mailing lists, and Twitter. We collected survey responses for approximately a month, with a survey deadline of August 31, 2018. We told participants that we would be releasing the responses from this survey as a public resource (with the exception of e-mail addresses, which would be kept private). People contacted were free to decline participation in the survey. Only adult scientists were allowed to participate. The research was approved by the UPenn IRB (Protocol number 830156).

\begin{figure}[t]
\centering
\includegraphics[width=14cm]{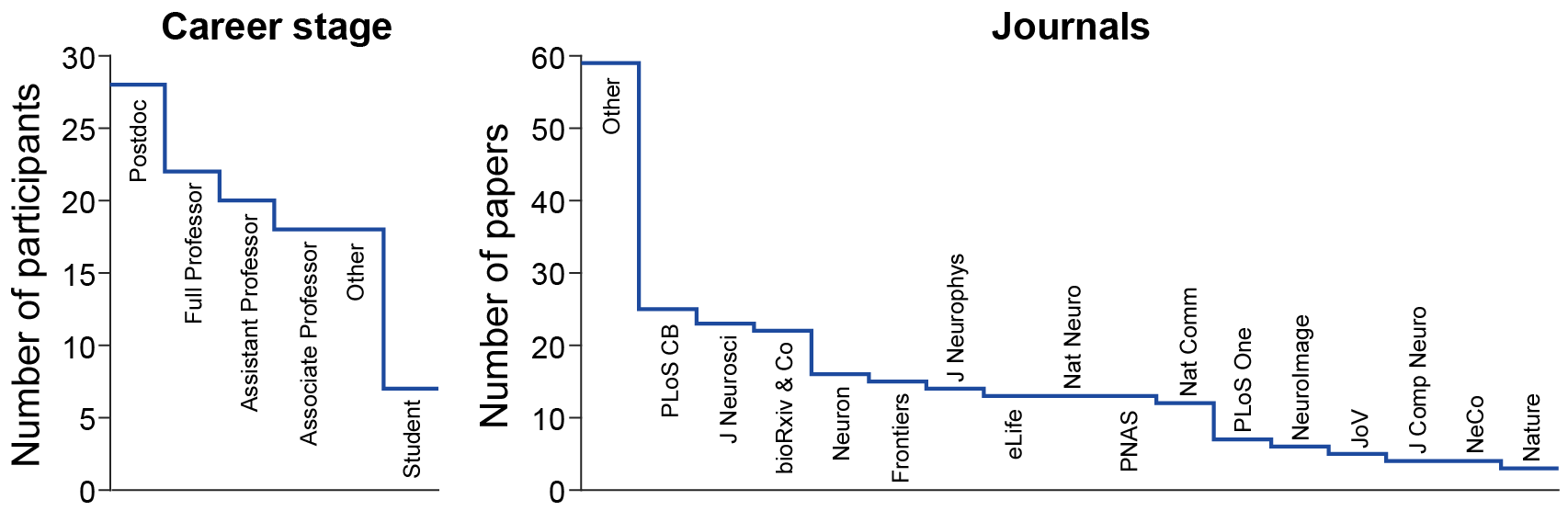}
\caption{Survey demographics. Shown are counts of survey participants binned by career stage (left) and counts of papers rated by survey participants binned by journal (right).}
\end{figure}

The survey asked each participant to choose up to 3 papers they authored or co-authored and to rate each paper on the 12 modeling goals described above.  Participants were instructed to submit papers representative of distinct types of their research. The full set of survey questions and survey results are available at \url{https://osf.io/pqe7f/}. We note that the survey results may be useful for answering a variety of additional questions not addressed in this paper. For example, one might be interested to compare one's prediction of the modeling goals held by a given researcher to the actual goals held by that researcher. Or as another example, one might be interested to see where one's goals fall relative to the group norms.

For the analyses performed in this paper, ratings were aggregated across papers (251 papers from 113 distinct authors; 22 female, 91 male). For Figure 2B, a small amount of Gaussian noise (mean 0, standard deviation 0.5) was added to the data prior to computing summary statistics in order to avoid discretization effects.

\section{Results}

\subsection{A simple survey provides evidence computational researchers have diverse goals}

To empirically assess modeling goals, we conducted an informal online survey in which we asked authors to rate their own modeling work with respect to the goals listed above. Participants rated up to 3 of their authored papers on each of the 12 goals, indicating the importance of each goal. We obtained results from 113 distinct authors who rated a total of 251 papers (Figure 1). On average, interpretability was rated as the most important modeling goal, whereas clinical relevance was rated as least important (Figure 2B, black bars). In addition, we found large variance of ratings across papers (Figure 2B, gray error bars), suggesting that there is, indeed, wide diversity of modeling goals in the neuroscience community. Not surprisingly, some goals are highly correlated (Figure 2C), such as compactness and tractability.

To better understand the underlying structure of the ratings, we subtracted the average rating of each modeling goal and identified a lower-dimensional space using probabilistic principal components analysis (Figure 3A). We reconstructed the data in this lower-dimensional space and recomputed the pairwise correlation structure (Figure 3B). Finally, we re-ordered the modeling goals, revealing three groups or clusters (Figure 3C). One simple interpretation of these clusters is that people are sampling independently mixed contributions from three clusters that somewhat overlap with the intuitive grouping of \emph{Scientific impact}, \emph{Biological realism}, and \emph{Style} (as shown in Figure 2A). These differences might represent different subfields of neuroscience having different modeling goals and/or different types of models naturally fulfilling certain goals more readily than others. However, this lower-dimensional reconstruction accounted for only 51\% of the total variance in goal ratings. The remaining 49\% of the variance reflects diversity of individual preferences in goals. Thus, just like the contrast between Gunnar's and Konrad's linear-systems models, the variability in modeling goals between researchers appears to be high.

\begin{figure}[t]
\centering
\includegraphics[width=14cm]{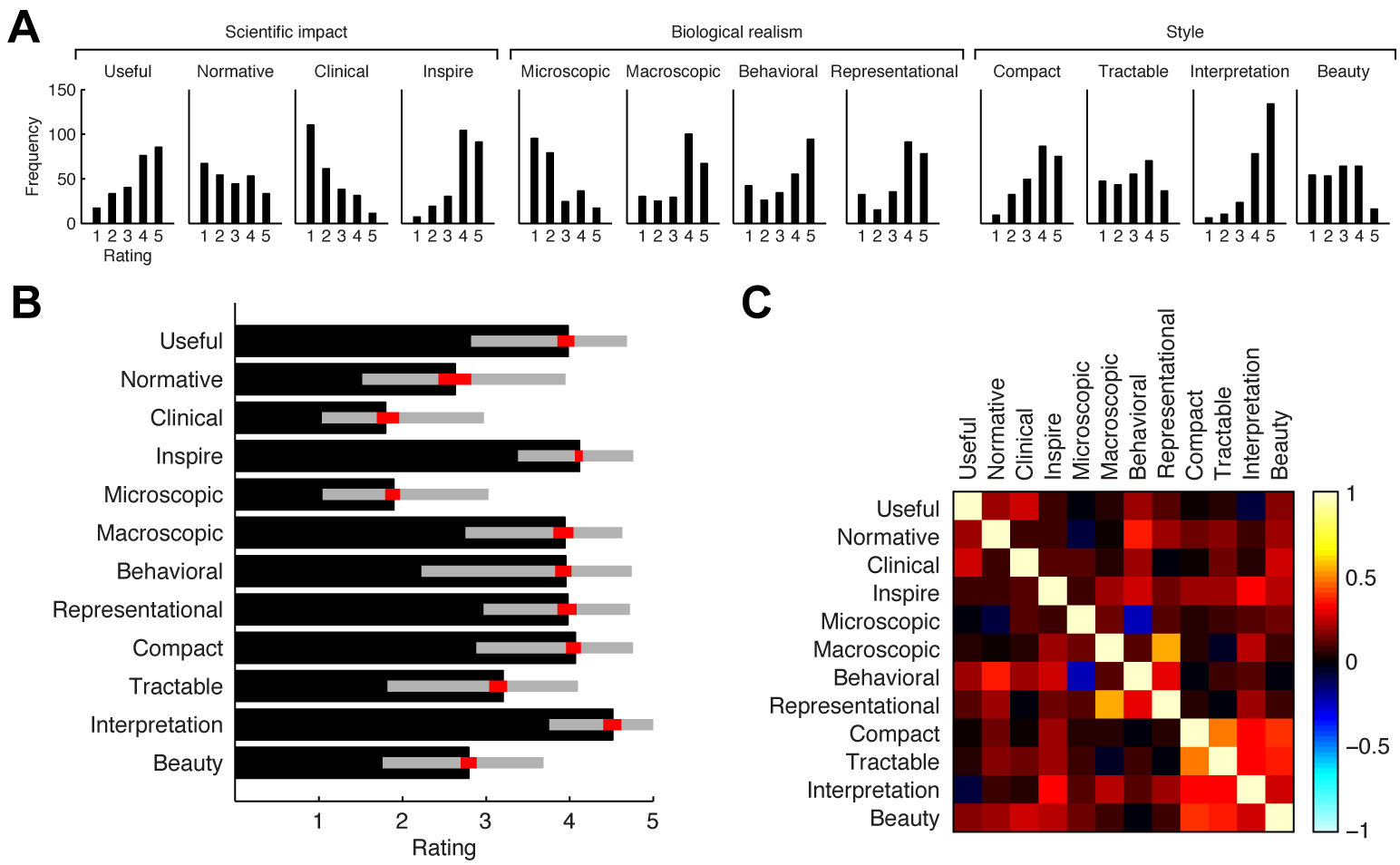}
\caption{Modeling goals in the computational neuroscience community. We conducted an informal survey on the Internet to assess the modeling goals that different researchers held for specific papers that they authored. (A) Histogram of results. For each of 12 modeling goals (dimensions), we plot a histogram of the reported ratings. Values range from 1 (completely irrelevant) to 5 (absolutely essential). Dimensions are ordered according to three conceptual groups that seem intuitively reasonable. (B) Summary statistics. For each dimension, we plot the median (black bars), interquartile range (gray error bars), and bootstrapped 68\% confidence interval on the median (red error bars). (C) Pairwise correlation (Pearson's \emph{r}) of dimensions across all papers.}
\end{figure}To further illustrate diversity, we visualize the location of each paper in the space spanned by the first three principal components (Figure 4). This shows again that despite dimensionality reduction accounting for roughly half of the variance, there are no discernible clusters or groups in this subspace. In other words, there is a continuum of preferences with respect to modeling goals that span the space. Highlighting the authors' own papers in Figure 4 allows several additional observations: (1) Most of the authors' papers' goals are fairly polarized, i.e. they reside at the edges of the space spanned. (2) Konrad's and Gunnar's papers lie in diametrically opposite sides of this space despite some overlap in modeling techniques used. The same is true for Paul's and Kendrick's papers. (3) Somewhat surprisingly, Paul's and Gunnar's goals (and Kendrick's and Konrad's goals) align fairly well despite apparently different technical approaches used. Note that this diversity in goals and approaches does not mean that we do not appreciate each other's research efforts; quite the opposite! We view this diversity as a strength (see Discussion).

\subsection{Limitations of the survey}

Of course, this survey is not intended as a formal scientific instrument to identify modeling goals in the field, but the results do suggest that it would be worthwhile to invest in more systematic meta-scientific analyses. It would be valuable to identify modeling goals for a much larger sample of papers and to calibrate the survey and analysis methods. Our sample is small and not randomly drawn from the population. The survey questions themselves may not have been understood in exactly the same way by different participants, which may have increased the apparent diversity. But the survey is sufficient to make our point: diversity in modeling goals is real and high! 

\begin{figure}[t]
\centering
\includegraphics[width=14cm]{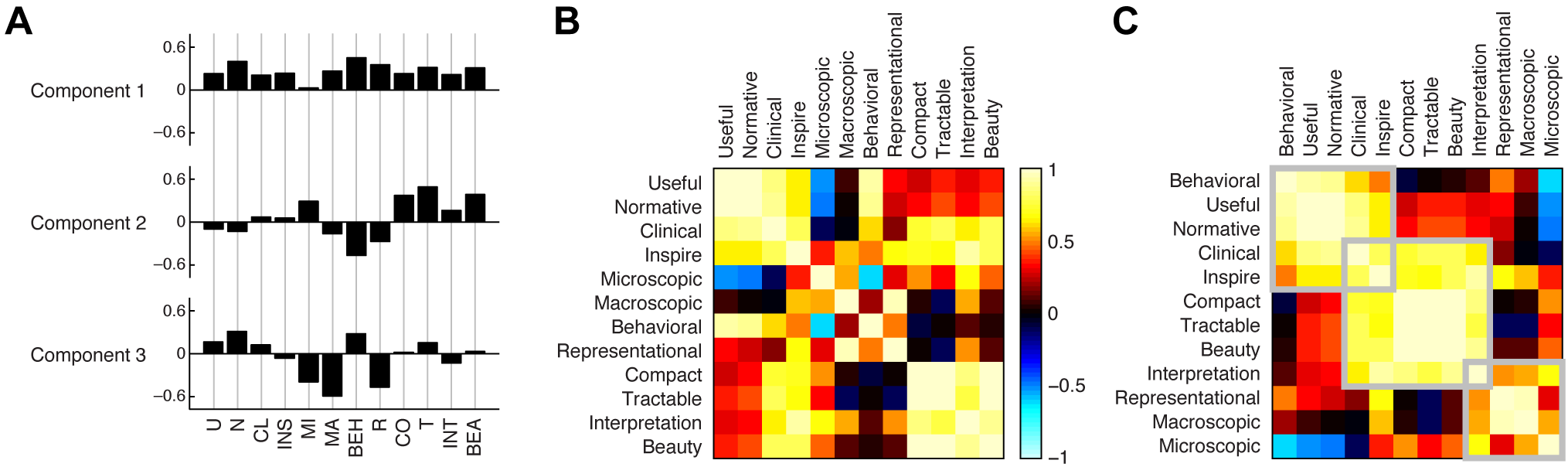}
\caption{Groups of modeling goals. We performed probabilistic principal component analysis to characterize the latent covariance structure of the survey ratings. (A) Component loadings. The top three components (composing an orthonormal basis) are shown; the order of dimensions is the same as in Figure 2. (B) Data reconstruction. We reconstruct the original data using the three identified components and recompute the pairwise correlation as in Figure 2C. The correlation structure is similar to that of the original data, but stronger due to the dimensionality reduction. (C) Grouping of dimensions. We re-plot the results of panel B, re-ordering dimensions to highlight the block structure. Thick gray squares indicate three groups of dimensions that appear to be present in the data.}
\end{figure}

\section{Discussion}

\subsection{Goals matter to how modeling is done}

The choice of goals matters in just about every imaginable way for modeling in neuroscience. Modeling goals affect the overall utility and interpretation of a model by influencing the evaluation metrics, the choice of model type, and the way we replace models with newer, better models. For example, if microscopic realism is required, this severely constrains the types of modeling techniques that can be used. Some research fields have implicit agreements on a set of desirable modeling criteria. For instance, historically, the eye-movements field has used linear-systems theory to model saccades: the field has been most concerned with behavioral realism, usefulness, inspiration of experiments, and interpretability, but has not placed much value on microscopic realism. If the field had been concerned with microscopic realism, the linear-systems approach would have likely been inappropriate and a different toolset—such as spiking neural-network models—would have been used instead. Thus, getting clarity on \emph{why} we model may be just as important as understanding the mechanics of \emph{how} to model.

\begin{figure}[t]
\centering
\includegraphics[width=14cm]{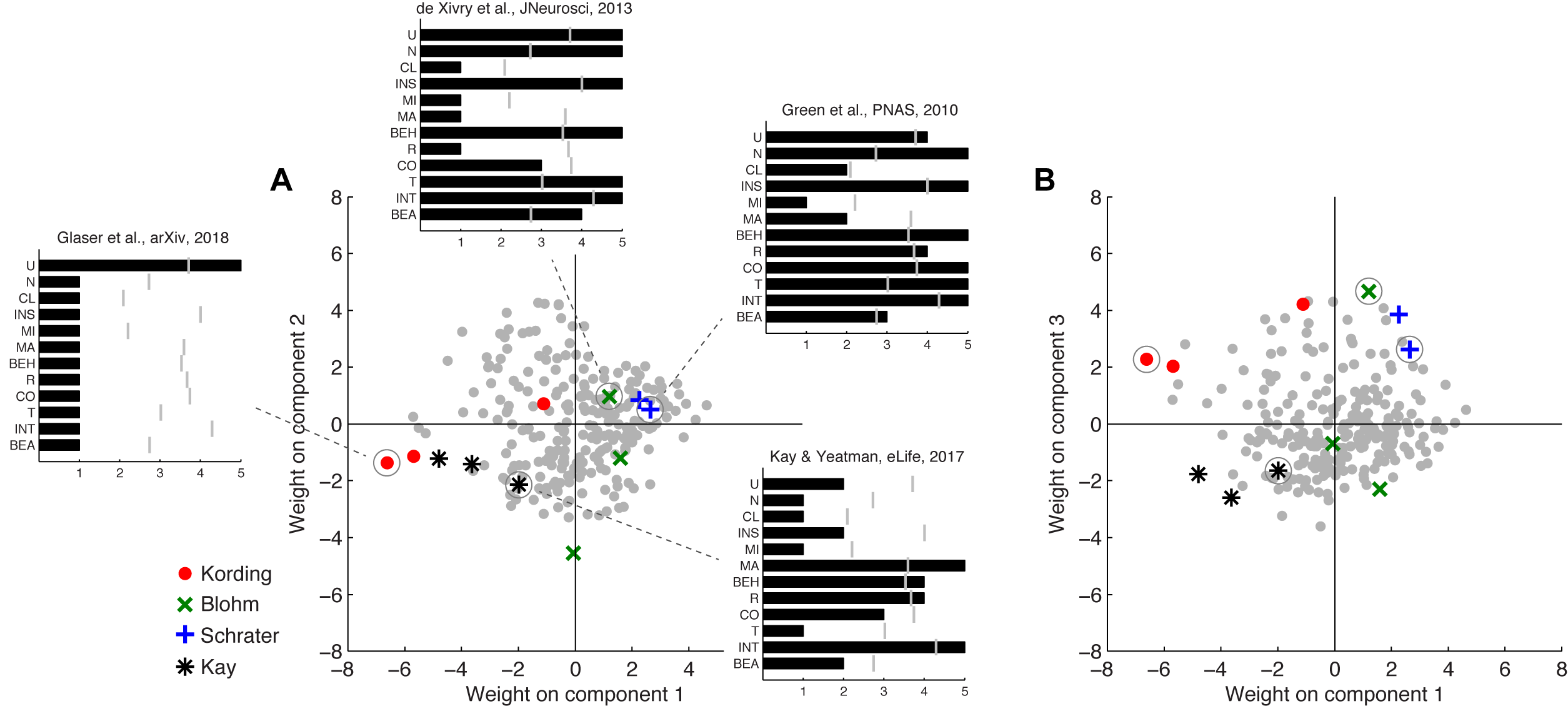}
\caption{Diversity of modeling goals within and across authors. Here we plot all recorded paper ratings in the space defined by the first three principal components (see Figure 3A). (A) Space defined by components 1 and 2. Each gray dot indicates a single paper, and colored markers indicate papers from the authors. Insets show raw ratings for example papers; in these plots, thin gray lines indicate the mean of each dimension across all papers. Some authors are consistent across papers they write (Schrater, Kay), whereas other authors show more diversity in their papers (Kording, Blohm). Furthermore, there is high diversity in modeling goals across the four authors of this paper. (B) Space defined by components 1 and 3. Same format as panel A.}
\end{figure}

\subsection{But what are the goals of computational neuroscience?}

In shaping our list of goals of computational neuroscience, we drew from our own experience: this is thus an opinion piece in which we try to provide a meaningful perspective to the field. We combine 7+ years of teaching students the broad range of modeling techniques useful in movement science and extensively discussing modeling objectives at the Summer School in Computational Sensory-Motor Neuroscience (CoSMo, \url{http://www.compneurosci.com/CoSMo/}) and beyond. That being said, while goals in neuroscience are diverse as we have shown, we cannot claim that our list is exhaustive or that other neuroscientists would not structure it differently. Even if our views are wrong, we hope that our humble paper will jump-start a drive towards clarity in modeling goals in neuroscience.

\subsection{Authors should state goals; readers should evaluate based on those commitments}

Why does it matter that different researchers have different modeling goals? We wish to raise awareness because diversity often leads to significant tension and misunderstandings between researchers. For example, a reviewer might have a certain set of goals associated with a particular modeling approach and might evaluate a given paper outside of the authors' intentions. This out-of-scope evaluation is one of the most frequent and frustrating reasons for miscommunication in computational neuroscience. Its consequence is often paper rejection. We argue that this behavior is detrimental for science.

What are some practical steps we can take? As action items, we suggest that (1) authors should explicitly spell out their goals, ideally in multiple instances across the Introduction, Methods, and Discussion sections \citep{blohmHowtomodelGuideNeuroscience2018,schraterModelingNeuroscienceDecision2019}, and (2) readers should deliberately evaluate a given paper within those constraints. For example, an author might include a section that begins, ``In this paper, we sought to satisfy the criteria of X, Y, and Z. It is not our goal to develop a model that exhibits properties A, B, and C for the following reasons...'' We believe that explicit characterization of goals leads to more constructive interactions and therefore promotes scientific progress, discovery, and societal impact.

This paper is itself a (rudimentary) modeling effort: we sought to characterize the intentions or goals held by computational neuroscientists in conducting their research and the relationship of these goals to one another. Thus, in a sense, this is behavioral research on how brains (scientists) function. With respect to the 12 modeling criteria, we sought to characterize the phenomenon at the behavioral level, with some interest in identifying the underlying representations (latent structure). Interpretability of our results was paramount, and we sought to gather observations that may inspire further meta-scientific efforts. On the other hand, our investigation is obviously not intended to understand the macroscopic or microscopic neural mechanisms underlying how scientists conduct their research. Likewise, several modeling criteria are clearly not applicable to our study (e.g. normative, clinical).

\subsection{Diversity of modeling goals is a strength}

Modeling aims to generate insight into a phenomenon of interest. Since models in computational neuroscience all refer to brains, one could argue that they are guaranteed to produce synergistic answers and that the distinctions highlighted in this paper are not that important. We think this stance is debatable. Models positioned at different levels of biological realism (microscopic, macroscopic, behavioral, representational) are not guaranteed to inform each other, as distinct phenomena may emerge at different levels. Models following different styles (compact, tractable, interpretation, beauty) are optimizing different criteria, so a model that fulfills one criterion may be suboptimal under other criteria. Models aimed towards a specific type of scientific impact (useful, normative, clinical, inspire) often fail to deliver other types of impact. Thus, it is our contention that modeling goals are truly diverse and that models in computational neuroscience are not aimed towards a single coherent class of answers.

Should the community attempt to converge on a single set of standards? While this might seem appealing to some, actual diversity of modeling goals makes it difficult to find a shared community preference function. From Arrow's Impossibility Theorem \citep{fishburnArrowImpossibilityTheorem1970} to Harsanyi's Aggregation Theorem \citep{fleurbaeyTwoVariantsHarsanyi2009,harsanyiBayesianDecisionTheory1979,weymarkHarsanyiSocialAggregation1993}, arriving at a consistent group preference function is known to be hard, to require special conditions, and even when possible, to require trading off individual preferences to allow a non-unique group preference as a point on a Pareto front. Editor and reviewer preferences for goals are unlikely to represent such a hypothetical aggregate preference, thus leading to idiosyncratic critiques \citep{garfunkelProblemsIdentifiedSecondary1990}. Rather than encourage conformity to particular preferences, appreciating goal diversity allows exploring possibilities without getting stuck in local minima. For example, 15 years ago, some researchers were claiming that working on machine learning was career suicide; researchers that nonetheless persevered are now superstars in the field.

We advocate embracing diversity in modeling goals as a strength for the field. As in other aspects of life, humanity works best by respecting and not excluding diversity. In terms of scientific progress, diversity balances biases, provides alternative views, encourages discussion, invigorates problem-solving, and facilitates specialization of individual researchers, each of whom can make distinct meaningful contributions to the field. Perhaps one day the neuroscience community will come to consensus on a single framework for describing and understanding the brain. But until that day comes, embracing diversity and explicitly recognizing each other's modeling goals will be critical for achieving progress.



%


%

\section*{acknowledgements}
We would like to thank the reviewers and Editor J. Pillow for constructive feedback.

\section*{conflict of interest}
The authors declare no conflicts of interest.






\begin{thebibliography}{52}
\expandafter\ifx\csname natexlab\endcsname\relax\def\natexlab#1{#1}\fi
\expandafter\ifx\csname url\endcsname\relax
  \def\url#1{\texttt{#1}}\fi
\expandafter\ifx\csname urlprefix\endcsname\relax\def\urlprefix{URL: }\fi

\bibitem[{Acu{\~n}a and Schrater(2010)}]{acunaStructureLearningHuman2010}
Acu{\~n}a, D.~E. and Schrater, P. (2010) Structure learning in human sequential
  decision-making.
\newblock \textit{PLoS computational biology}, \textbf{6}, e1001003.

\bibitem[{{Al-Nashash} et~al.(2004){Al-Nashash}, {Al-Assaf}, Paul and
  Thakor}]{al-nashashEEGSignalModeling2004}
{Al-Nashash}, H., {Al-Assaf}, Y., Paul, J. and Thakor, N. (2004) {{EEG}} signal
  modeling using adaptive {{Markov}} process amplitude.
\newblock \textit{IEEE transactions on bio-medical engineering}, \textbf{51},
  744--751.

\bibitem[{Bai et~al.(2019)Bai, Martin, Guo, Dokos and
  Loo}]{baiComputationalComparisonConventional2019}
Bai, S., Martin, D., Guo, T., Dokos, S. and Loo, C. (2019) Computational
  comparison of conventional and novel electroconvulsive therapy electrode
  placements for the treatment of depression.
\newblock \textit{European Psychiatry: The Journal of the Association of
  European Psychiatrists}, \textbf{60}, 71--78.

\bibitem[{Barlow(1961)}]{barlowPossiblePrinciplesUnderlying1961}
Barlow, H.~B. (1961) Possible principles underlying the transformation of
  sensory messages.
\newblock \textit{Sensory communication}, \textbf{1}, 217--234.

\bibitem[{Bennett and
  Hacker(2003)}]{bennettPhilosophicalFoundationsNeuroscience2003}
Bennett, M.~R. and Hacker, P. M.~S. (2003) \textit{Philosophical Foundations of
  Neuroscience}, vol.~79.
\newblock {Blackwell Oxford}.

\bibitem[{Blohm et~al.(2009)Blohm, Keith and
  Crawford}]{blohmDecodingCorticalTransformations2009}
Blohm, G., Keith, G.~P. and Crawford, J.~D. (2009) Decoding the cortical
  transformations for visually guided reaching in {{3D}} space.
\newblock \textit{Cerebral Cortex}, \textbf{19}, 1372--1393.

\bibitem[{Blohm et~al.(2018)Blohm, Kording and
  Schrater}]{blohmHowtomodelGuideNeuroscience2018}
Blohm, G., Kording, K.~P. and Schrater, P.~R. (2018) A how-to-model guide for
  {{Neuroscience}}.
\newblock \textit{OSF Preprints}.

\bibitem[{Bornmann et~al.(2010)Bornmann, Weymuth and
  Daniel}]{bornmannContentAnalysisReferees2010}
Bornmann, L., Weymuth, C. and Daniel, H.-D. (2010) A content analysis of
  referees' comments: How do comments on manuscripts rejected by a high-impact
  journal and later published in either a low-or high-impact journal differ?
\newblock \textit{Scientometrics}, \textbf{83}, 493--506.

\bibitem[{Burgess(1998)}]{burgessOccamRazorScientific1998}
Burgess, J. (1998) Occam's razor and scientific method.
\newblock In \textit{The {{Philosophy}} of {{Mathematics Today}}}, 195--214.
  {Clarendon Press Oxford}.

\bibitem[{Byrne(2000)}]{byrneCommonReasonsRejecting2000}
Byrne, D.~W. (2000) Common reasons for rejecting manuscripts at medical
  journals: {{A}} survey of editors and peer reviewers.
\newblock \textit{Science Editor}.

\bibitem[{Carlson et~al.(2003)Carlson, Schrater and
  He}]{carlsonPatternsActivityCategorical2003a}
Carlson, T.~A., Schrater, P. and He, S. (2003) Patterns of activity in the
  categorical representations of objects.
\newblock \textit{Journal of Cognitive Neuroscience}, \textbf{15}, 704--717.

\bibitem[{Chandrasekhar(2013)}]{chandrasekharTruthBeautyAesthetics2013}
Chandrasekhar, S. (2013) \textit{Truth and {{Beauty}}: {{Aesthetics}} and
  {{Motivations}} in {{Science}}}.
\newblock {University of Chicago Press}.

\bibitem[{Chater and Oaksford(1999)}]{chaterTenYearsRational1999}
Chater, N. and Oaksford, M. (1999) Ten years of the rational analysis of
  cognition.
\newblock \textit{Trends in Cognitive Sciences}, \textbf{3}, 57--65.

\bibitem[{Chater and Oaksford(2000)}]{chaterRationalAnalysisMind2000}
--- (2000) The rational analysis of mind and behavior.
\newblock \textit{Synthese}, \textbf{122}, 93--131.

\bibitem[{Churchland and
  Sejnowski(1990)}]{churchlandNeuralRepresentationNeural1990}
Churchland, P.~S. and Sejnowski, T.~J. (1990) Neural representation and neural
  computation.
\newblock \textit{Philosophical Perspectives}, \textbf{4}, 343--382.

\bibitem[{Dan and Poo(2004)}]{danSpikeTimingdependentPlasticity2004}
Dan, Y. and Poo, M.-M. (2004) Spike timing-dependent plasticity of neural
  circuits.
\newblock \textit{Neuron}, \textbf{44}, 23--30.

\bibitem[{Dayan and
  Abbott(2001)}]{dayanTheoreticalNeuroscienceComputational2001}
Dayan, P. and Abbott, L.~F. (2001) \textit{Theoretical Neuroscience:
  Computational and Mathematical Modeling of Neural Systems}.
\newblock {MIT Press}.

\bibitem[{Fishburn(1970)}]{fishburnArrowImpossibilityTheorem1970}
Fishburn, P.~C. (1970) Arrow's impossibility theorem: {{Concise}} proof and
  infinite voters.
\newblock \textit{Journal of Economic Theory}, \textbf{2}, 103--106.

\bibitem[{Fitts and Radford(1966)}]{fittsInformationCapacityDiscrete1966}
Fitts, P.~M. and Radford, B.~K. (1966) Information capacity of discrete motor
  responses under different cognitive sets.
\newblock \textit{Journal of Experimental Psychology}, \textbf{71}, 475--482.

\bibitem[{Fleurbaey(2009)}]{fleurbaeyTwoVariantsHarsanyi2009}
Fleurbaey, M. (2009) Two variants of {{Harsanyi}}'s aggregation theorem.
\newblock \textit{Economics Letters}, \textbf{105}, 300--302.

\bibitem[{Fukushima(1980)}]{fukushimaNeocognitronSelfOrganizing1980a}
Fukushima, K. (1980) Neocognitron: A self organizing neural network model for a
  mechanism of pattern recognition unaffected by shift in position.
\newblock \textit{Biological Cybernetics}, \textbf{36}, 193--202.

\bibitem[{Garfunkel et~al.(1990)Garfunkel, Ulshen, Hamrick and
  Lawson}]{garfunkelProblemsIdentifiedSecondary1990}
Garfunkel, J.~M., Ulshen, M.~H., Hamrick, H.~J. and Lawson, E.~E. (1990)
  Problems identified by secondary review of accepted manuscripts.
\newblock \textit{JAMA}, \textbf{263}, 1369--1371.

\bibitem[{Gerstner et~al.(1996)Gerstner, Kempter, {van Hemmen} and
  Wagner}]{gerstnerNeuronalLearningRule1996}
Gerstner, W., Kempter, R., {van Hemmen}, J.~L. and Wagner, H. (1996) A neuronal
  learning rule for sub-millisecond temporal coding.
\newblock \textit{Nature}, \textbf{383}, 76--81.

\bibitem[{Gillett(2016)}]{gillettReductionEmergenceScience2016}
Gillett, C. (2016) \textit{Reduction and Emergence in Science and Philosophy}.
\newblock {Cambridge University Press}.

\bibitem[{Green et~al.(2010)Green, Benson, Kersten and
  Schrater}]{greenAlterationsChoiceBehavior2010}
Green, C.~S., Benson, C., Kersten, D. and Schrater, P. (2010) Alterations in
  choice behavior by manipulations of world model.
\newblock \textit{Proceedings of the National Academy of Sciences of the United
  States of America}, \textbf{107}, 16401--16406.

\bibitem[{Harris and Wolpert(1998)}]{harrisSignaldependentNoiseDetermines1998}
Harris, C.~M. and Wolpert, D.~M. (1998) Signal-dependent noise determines motor
  planning.
\newblock \textit{Nature}, \textbf{394}, 780--784.

\bibitem[{Harsanyi(1979)}]{harsanyiBayesianDecisionTheory1979}
Harsanyi, J.~C. (1979) Bayesian decision theory, rule utilitarianism, and
  {{Arrow}}'s impossibility theorem.
\newblock \textit{Theory and Decision}, \textbf{11}, 289--317.

\bibitem[{Josephson and
  Josephson(1996)}]{josephsonAbductiveInferenceComputation1996}
Josephson, J.~R. and Josephson, S.~G. (1996) \textit{Abductive {{Inference}}:
  {{Computation}}, {{Philosophy}}, {{Technology}}}.
\newblock {Cambridge University Press}.

\bibitem[{Kay and Yeatman(2017)}]{kayBottomupTopdownComputations2017a}
Kay, K.~N. and Yeatman, J.~D. (2017) Bottom-up and top-down computations in
  word- and face-selective cortex.
\newblock \textit{eLife}, \textbf{6}.

\bibitem[{Knill and Richards(1996)}]{knillPerceptionBayesianInference1996}
Knill, D.~C. and Richards, W. (1996) \textit{Perception as {{Bayesian}}
  Inference}.
\newblock {Cambridge University Press}.

\bibitem[{K{\"o}rding(2007)}]{kordingDecisionTheoryWhat2007}
K{\"o}rding, K. (2007) Decision theory: What "should" the nervous system do?
\newblock \textit{Science}, \textbf{318}, 606--610.

\bibitem[{Kording et~al.(2007)Kording, Tenenbaum and
  Shadmehr}]{kordingDynamicsMemoryConsequence2007}
Kording, K.~P., Tenenbaum, J.~B. and Shadmehr, R. (2007) The dynamics of memory
  as a consequence of optimal adaptation to a changing body.
\newblock \textit{Nature Neuroscience}, \textbf{10}, 779--786.

\bibitem[{Krakauer et~al.(2017)Krakauer, Ghazanfar, {Gomez-Marin}, MacIver and
  Poeppel}]{krakauerNeuroscienceNeedsBehavior2017}
Krakauer, J.~W., Ghazanfar, A.~A., {Gomez-Marin}, A., MacIver, M.~A. and
  Poeppel, D. (2017) Neuroscience {{Needs Behavior}}: {{Correcting}} a
  {{Reductionist Bias}}.
\newblock \textit{Neuron}, \textbf{93}, 480--490.

\bibitem[{Li and Vit{\'a}nyi(2019)}]{liIntroductionKolmogorovComplexity2019}
Li, M. and Vit{\'a}nyi, P. (2019) \textit{An Introduction to {{Kolmogorov}}
  Complexity and Its Applications}, vol.~3.
\newblock {Springer}.

\bibitem[{Lombrozo(2012)}]{lombrozoExplanationAbductiveInference2012}
Lombrozo, T. (2012) Explanation and abductive inference.
\newblock In \textit{Oxford Handbook of Thinking and Reasoning}, 260--276.
  {Oxford University Press}.

\bibitem[{Ma et~al.(2006)Ma, Beck, Latham and
  Pouget}]{maBayesianInferenceProbabilistic2006a}
Ma, W.~J., Beck, J.~M., Latham, P.~E. and Pouget, A. (2006) Bayesian inference
  with probabilistic population codes.
\newblock \textit{Nature Neuroscience}, \textbf{9}, 1432--1438.

\bibitem[{Mayr(2004)}]{mayrWhatMakesBiology2004}
Mayr, E. (2004) \textit{What Makes Biology Unique?: Considerations on the
  Autonomy of a Scientific Discipline}.
\newblock {Cambridge University Press}.

\bibitem[{Mukherjee(2018)}]{mukherjee11ReasonsWhy2018}
Mukherjee, D. (2018) 11 {{Reasons Why Research Papers Are Rejected}}.
\newblock
  https://blog.typeset.io/11-reasons-why-research-papers-are-rejected-3e272b633186.

\bibitem[{Olshausen and Field(2004)}]{olshausenSparseCodingSensory2004}
Olshausen, B.~A. and Field, D.~J. (2004) Sparse coding of sensory inputs.
\newblock \textit{Current opinion in neurobiology}, \textbf{14}, 481--487.

\bibitem[{{Orban de Xivry} et~al.(2013){Orban de Xivry}, Coppe, Blohm and
  Lef{\`e}vre}]{orbandexivryKalmanFilteringNaturally2013}
{Orban de Xivry}, J.-J., Coppe, S., Blohm, G. and Lef{\`e}vre, P. (2013) Kalman
  filtering naturally accounts for visually guided and predictive smooth
  pursuit dynamics.
\newblock \textit{The Journal of Neuroscience}, \textbf{33}, 17301--17313.

\bibitem[{Parker(2012)}]{parkerComputerSimulationPhilosophy2012}
Parker, W.~S. (2012) Computer simulation and philosophy of science.
\newblock \textit{Metascience}, \textbf{21}, 111--114.

\bibitem[{Pierson(2004)}]{piersonTop10Reasons2004}
Pierson, D.~J. (2004) The top 10 reasons why manuscripts are not accepted for
  publication.
\newblock \textit{Respiratory Care}, \textbf{49}, 1246--1252.

\bibitem[{Russell(2019)}]{russellMysticismLogicOther2019}
Russell, B. (2019) \textit{Mysticism and {{Logic}} and {{Other Essays}}}.
\newblock {Good Press}.

\bibitem[{Schrater et~al.(2019)Schrater, Kording and
  Blohm}]{schraterModelingNeuroscienceDecision2019}
Schrater, P., Kording, K. and Blohm, G. (2019) Modeling in {{Neuroscience}} as
  a {{Decision Process}}.
\newblock \textit{OSF Preprints}.

\bibitem[{Selinger et~al.(2015)Selinger, O'Connor, Wong and
  Donelan}]{selingerHumansCanContinuously2015}
Selinger, J.~C., O'Connor, S.~M., Wong, J.~D. and Donelan, J.~M. (2015) Humans
  {{Can Continuously Optimize Energetic Cost}} during {{Walking}}.
\newblock \textit{Current biology}, \textbf{25}, 2452--2456.

\bibitem[{Serre and Riesenhuber(2004)}]{serreRealisticModelingSimple2004}
Serre, T. and Riesenhuber, M. (2004) Realistic {{Modeling}} of {{Simple}} and
  {{Complex Cell Tuning}} in the {{HMAXModel}}, and {{Implications}} for
  {{Invariant Object Recognition}} in {{Cortex}}.
\newblock \textit{Tech. rep.}, {Massachusetts Institute of Technology}.

\bibitem[{Tao et~al.(2018)Tao, Khan and Blohm}]{taoCorrectiveResponseTimes2018}
Tao, G., Khan, A.~Z. and Blohm, G. (2018) Corrective response times in a
  coordinated eye-head-arm countermanding task.
\newblock \textit{Journal of Neurophysiology}, \textbf{119}, 2036--2051.

\bibitem[{Thrower(2012)}]{throwerEightReasonsRejected2012}
Thrower, P. (2012) Eight reasons {{I}} rejected your article.
\newblock https://www.elsevier.com/connect/8-reasons-i-rejected-your-article.

\bibitem[{Todorov and Jordan(2002)}]{todorovOptimalFeedbackControl2002}
Todorov, E. and Jordan, M.~I. (2002) Optimal feedback control as a theory of
  motor coordination.
\newblock \textit{Nature Neuroscience}, \textbf{5}, 1226--1235.

\bibitem[{Webster(2016)}]{websterMerriamWebsterDictionaryInternational2016}
Webster (2016) \textit{The {{Merriam}}-{{Webster Dictionary}}: {{International
  Edition}}}.
\newblock {Merriam-Webster}.

\bibitem[{Weymark(1993)}]{weymarkHarsanyiSocialAggregation1993}
Weymark, J.~A. (1993) Harsanyi's social aggregation theorem and the weak
  {{Pareto}} principle.
\newblock \textit{Social choice and welfare}, \textbf{10}, 209--221.

\bibitem[{Zador et~al.(1990)Zador, Koch and
  Brown}]{zadorBiophysicalModelHebbian1990}
Zador, A., Koch, C. and Brown, T.~H. (1990) Biophysical model of a {{Hebbian}}
  synapse.
\newblock \textit{Proceedings of the National Academy of Sciences},
  \textbf{87}, 6718--6722.

\end{thebibliography}
\end{document}